\documentclass[aps,pre,reprint,showpacs,floatfix,groupedaddress,nofootinbib]{revtex4-1}

\usepackage{amscd}
\usepackage{amsmath}
\usepackage{amssymb}
\usepackage{amsthm}
\usepackage{bm}
\usepackage{epsfig}
\usepackage{epstopdf}
\usepackage{graphicx}
\usepackage{latexsym}
\usepackage{mathtools}
\usepackage{mathrsfs}
\usepackage{subfigure}
\usepackage{appendix}
\usepackage{multirow}
\usepackage{tikz} 
\usetikzlibrary{positioning}

\makeatletter
\DeclareRobustCommand*\textsubscript[1]{%
  \@textsubscript{\selectfont#1}}
\def\@textsubscript#1{%
  {\m@th\ensuremath{_{\mbox{\fontsize\sf@size\z@#1}}}}}
\makeatother

\begin{document}

\title{Weyl Rings and enhanced susceptibilities in Pyrochlore Iridates: $k\cdot p$ Analysis of Cluster Dynamical Mean-Field Theory Results}
\author{Runzhi Wang$^{(1)}$, Ara Go$^{(2)}$ and Andrew Millis$^{(1)}$}
\affiliation{$^{(1)}$Department of Physics, Columbia University, New York, New York 10027} 
\affiliation{$^{(2)}$Center for Theoretical Physics of Complex Systems, Institute for Basic Science (IBS), Daejeon 34051, Republic of Korea}
\date{\today}

\begin{abstract}
We match analytic results to numerical calculations  to provide a detailed picture of the metal-insulator and topological transitions found in density functional plus cluster dynamical mean-field calculations of pyrochlore iridates.  We discuss the transition from Weyl metal to Weyl semimetal regimes, and then analyse in detail the properties of the Weyl semimetal phase and its evolution  into the topologically trivial insulator.  The energy scales in the Weyl semimetal phase are found to be very small, as are the anisotropy parameters. The electronic structure  can to a good approximation be described as `Weyl rings' and one of the two branches that contributes to the Weyl bands is essentially flat, leading to  enhanced susceptibilities. The optical longitudinal and Hall conductivities are determined; the frequency dependence includes pronounced features that reveal the basic energy scales of the Weyl semimetal phase. 
\end{abstract}


\maketitle

\section{Introduction}

The Ir-based pyrochlore iridate compounds have been proposed as  promising potential hosts for topological phases \cite{Pesin2010}. In particular, Wan, Turner, Vishwanath and Savrasov \cite{dft} used density functional plus $U$ methods to identify a  Weyl semimetal regime in the  all-in/all-out (AIAO) antriferromagnetic phase of Y$_2$Ir$_2$O$_7$. Clear indirect \cite{AIAO-Nd,AIAO-Eu,AIAO-Y-Eu-Nd} and direct \cite{Determination-AIAO-Nd2Ir2O7} experimental evidence of AIAO ordering has been reported  in the Nd, Y and Eu-based pyrochlore iridates but it is not clear whether the WSM phase exists in any of the experimentally studied compounds. Angle-resolved photoemission (ARPES) measurements reported the absence of Weyl points  in the Nd compound \cite{WSM-ARPES-Nd} but recent optical studies have been interpreted as providing at least indirect evidence of Weyl points in  the Eu and Sm compounds \cite{WSM-optical-Eu,Ueda16}.  

Following the original theoretical proposals \cite{Pesin2010,dft} the WSM phase was subsequently found in Hartree-Fock \cite{tb-HF1,tb-HF2} and cluster dynamical mean-field calculations \cite{AI} using  tight-binding models representative of the band structure. However as mentioned in Ref. \cite{review}, the relatively weak dispersion of the frontier orbitals implies that the Weyl semimetal phase may exist only in a narrow parameter regime.  On the other hand, density functional plus single-site dynamical mean-field (DFT+sDMFT) studies predicted a direct first-order transition from paramagnetic metal to topologically trivial AIAO insulator, without an intervening  WSM phase \cite{lda-dmft-Y,lda-dmft-R-dep}. 

We recently presented density functional plus cluster dynamical mean-field (DFT+CDMFT) calculations of the pyrochlore iridate compounds Lu$_2$Ir$_2$O$_7$, Y$_2$Ir$_2$O$_7$ and Eu$_2$Ir$_2$O$_7$ as a function of the intra-$d$ Ir interaction strength $U$ \cite{dft+cdmft}. For all of the compounds studied the same generic ground-state phase diagram was found, with three phases: a low $U$ paramagnetic metallic phase, an intermediate $U$ topologically nontrivial AIAO antiferromagnetic Weyl metal phase and a higher $U$ topologically trivial antiferromagnetic insulating phase. This qualitative difference between cluster and single-site DMFT is interesting because the corrections to the single-site approximation were expected to be relatively weak in electronically three dimensional materials. However in this previous work the issue of a Weyl semimetal phase was not discussed and the transition from the topologically nontrivial antiferromagnetic metal phase to the topologically trivial insulating phase was not analysed.

In this paper we  present a detailed analysis of the metal-insulator and topological transitions implied by the DFT+CDMFT calculations of Ref. \onlinecite{dft+cdmft}, focussing on the vicinity of the antiferromagnetic metal to antiferromagnetic insulator  transition. For definiteness, our analysis uses the band parameters derived for Y$_2$Ir$_2$O$_7$  despite the variation across compounds, but our previous work \cite{dft+cdmft} found that the compounds were very similar except for an over-all change of bandwidth, so we expect that the main conclusions apply to all the pyrochlore iridates.  The behavior in the vicinity of the transition is subtle, involving low energy scales and sensitive dependence on parameters; numerical issues related to the finite bath size in the dynamical mean-field solver mean that the results cannot simply be read off from the CDMFT results.  We employ analytical arguments based on the $k\cdot p$ theory of Ref.~\onlinecite{dft} to fit the results of the CDMFT calculations, obtaining a clear picture of the metal-insulator and topological transitions found within the CDMFT approximation. We find that the Weyl metal phase previously reported is separated from the trivial insulator by a Weyl semimetal phase with two unusual characteristics. First, the anisotropy is very weak, so that while in principle the low energy electronic structure is described by Weyl points, to a very good approximation one has a zero energy ring of states. Further, one of the two bands whose crossing produces the Weyl points is almost perfectly flat (almost no dependence of energy on momentum), leading to an interesting structure in the optical conductivity and an enhanced (but still not divergent) susceptibility. We expect the results may be useful in the ongoing interpretation of experimental data on the pyrochlore iridate materials. 

The rest of this paper is organized as follows. Section ~\ref{Overview} reviews  and extends our prior results, defines terminology and specifies the questions of interest here. Section ~\ref{kdotp} presents the basic formulas of $k\cdot p$ theory that will be used to analyse  our numerical data and their fit to the numerical data, and discusses the transitions between the Weyl metal, Weyl semimetal and insulator phases.   Section ~\ref{Results} presents the application of the $k\cdot p$ theory to our data  and response functions including the longitudinal and hall terms in the optical conductivities and the static polarizibilities relevant to the stability of the Weyl semimetal state.  Section ~\ref{Conclusion} is a summary and conclusion. An Appendix provides technical details of the calculation.

\section{WM-WSM transition \label{Overview}}
\begin{figure}
\begin{center}
\begin{tikzpicture}
\node (fig1) at (0,0)                                 { \includegraphics[width=0.99\linewidth ]{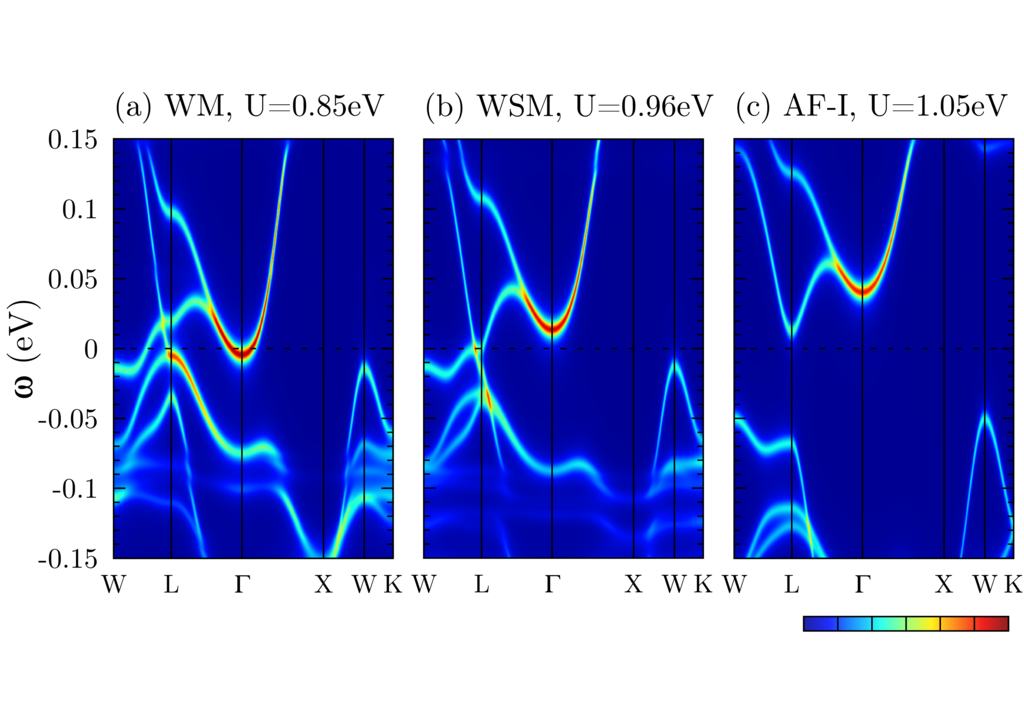} };
\node[anchor=north,xshift=4mm, yshift=20mm] (fig2) at (fig1.north) { \includegraphics[width=0.92\linewidth ]{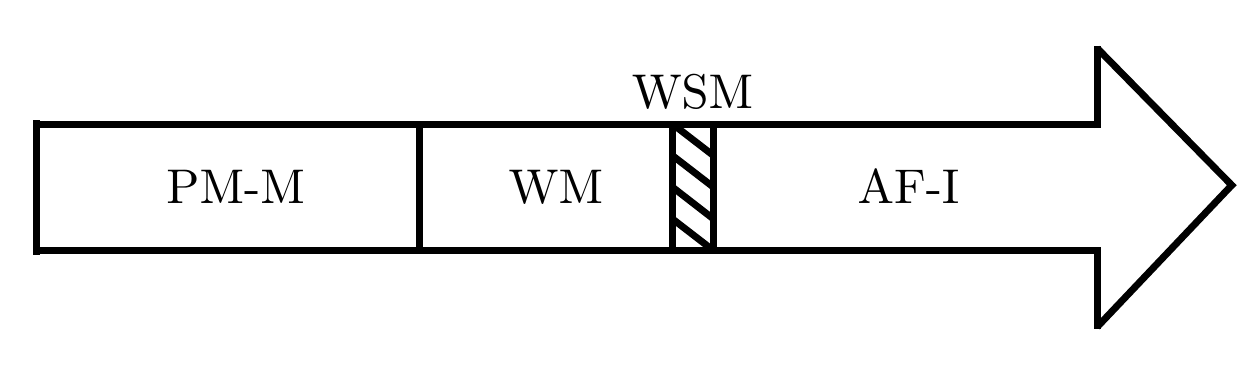} };
\end{tikzpicture}
\caption{Top panel: qualitative phase diagram of Y$_2$Ir$_2$O$_7$ as a function of on-site interaction strength $U$ (small $U$ at left, large $U$ at right) obtained by extending the DFT+CDMFT calculations of  Ref.~\onlinecite{dft+cdmft} to provide a more accurate treatment of the interaction strengths in the vicinity of the transition to the topologically trivial insulator, indicating paramagnetic metal (PM-M), antiferromagnetic Weyl metal (WM), Weyl semimetal (WSM; hatched) and topologically trivial antiferromagnetic insulator (AF-I) phases.  Lower panels: false-color representation of electron spectral function (Eq.~\ref{Adef}) as a function of frequncy ($y$-axis) for momenta along certain high-symmetry directions in the magnetic  Brillouin zone for (a) $U$=0.85eV (WM phase)  (b) $U$=0.96eV (WSM phase)  and (c) $U$=1.05eV (AF-I phase), with the zero of energy defined to be the Fermi level and broadening factor (Eq.~\ref{GRdef}) $\epsilon=0.005$eV. For all three U-values the states are magnetic with the AIAO magnetic ordering.}
\label{spec}
\end{center}
\end{figure}

In this section we extend and reinterpret our previous \cite{dft+cdmft} DFT+CDMFT results, in particular providing a more accurate treatment of the region of the transition to the topologically trivial insulating phase. The top panel of Fig. \ref{spec} shows the ground-state phase diagram found  for Y$_2$Ir$_2$O$_7$ as the correlation strength is varied. Weyl metal (WM) and Weyl semimetal (WSM) states separate a topologically trivial paramagnetic metal phase from an AIAO topologically trivial insulating phase (the WSM phase was not noted in our previous work). Qualitatively similar results are found  in DFT+$U$ \cite{dft} and tight-binding model-based Hartree-Fock \cite{tb-HF1,tb-HF2}  calculations. While the theoretical results are obtained by varying interaction strength at fixed composition, it is generally believed that varying the rare earth at fixed $U$ will produce a similar phase diagram, with paramagnetic Pr$_2$Ir$_2$O$_7$ representing the small-$U$ phase  and strongly insulating Lu$_2$Ir$_2$O$_7$  perhaps corresponding to the topologically trivial insulator. The  calculations predict a wide range of $U$ values for  which the material is an antiferromagnetic metal; it remains to be determined whether an antiferromagnetic metal phase is observed in any pyrochlore iridate.

We have computed the lattice Green's function (a matrix in the space of noninteracting bands); symmetrization of the CDMFT results is required; see the Appendix for technical details. We define the spectral function $A$ as  the trace of the branch cut discontinuity 

\begin{equation}
A(k,\omega)=-\frac{1}{2\pi i}Tr\left[\mathbf{G}^{(R)}(k,\omega)-\mathbf{G}^{(A)}(k,\omega)\right]
\label{Adef}
\end{equation}
with the retarded Green function
\begin{equation}
\mathbf{G}^{(R)}(k,\omega)=\left[\omega+i\epsilon+\mu-H_0(k)-\Sigma^{(R)}(k,\omega)\right]^{-1}
\label{GRdef}
\end{equation}
defined by letting the frequency approach the real axis from above and the advanced Green function defined by letting the frequency approach the real axis from below. $\epsilon$ is a broadening factor typically chosen to be $0.005eV$.

Quasiparticle bands are evident as regions where the spectral function is strongly enhanced.  Fig. \ref{spec} presents typical spectral functions for several values of the interaction strength in the magnetic phases.   Panel (a) shows results obtained for $U$=0.85eV. A band crossing point is apparent along the $W$-$L$ line, near to the $L$ point. This crossing point is protected by symmetry and we identify it as a Weyl crossing. The presence of a protected band crossing is a sign that the material is topologically non-trivial. A band dispersing upwards from the $\Gamma$ point is also observed. In panel (a) the band energy at the $\Gamma$ point is below the energy of the Weyl crossing point.  The total number of electrons per unit cell is even, so the electrons present in the band associated with the $\Gamma$ point must be compensated by holes in the bands below the Weyl crossing point; in other words,   the Fermi level must lie between the band energy at the $\Gamma$ point and the Weyl crossing energy. We therefore identify the $U$=0.85eV state as a Weyl metal.  

Comparison of panels (a) and (b) shows that as the interaction $U$ increases, the energy of the band minimum at the $\Gamma$ point increases relative to the energy of the Weyl crossing point; at some $U$-value the energy of the band minimum at $\Gamma$  becomes greater than the energy of the Weyl crossing. In this regime straightforward electron counting implies that the Fermi level must pass through the Weyl point. We identify this phase as the Weyl semimetal; it exists for a narrow range of $U$.  Panel (c) then shows that as the interaction strength is increased yet further, the Weyl crossing vanishes: the phase is a topologically trivial insulator. The key new result of this analysis is that the DFT+CDMFT method predicts a non-infinitesimal range of parameters over which the Weyl semimetal phase exists in the pyrochlore iridates.

\begin{figure}
\begin{center}
\includegraphics[width=1\columnwidth]{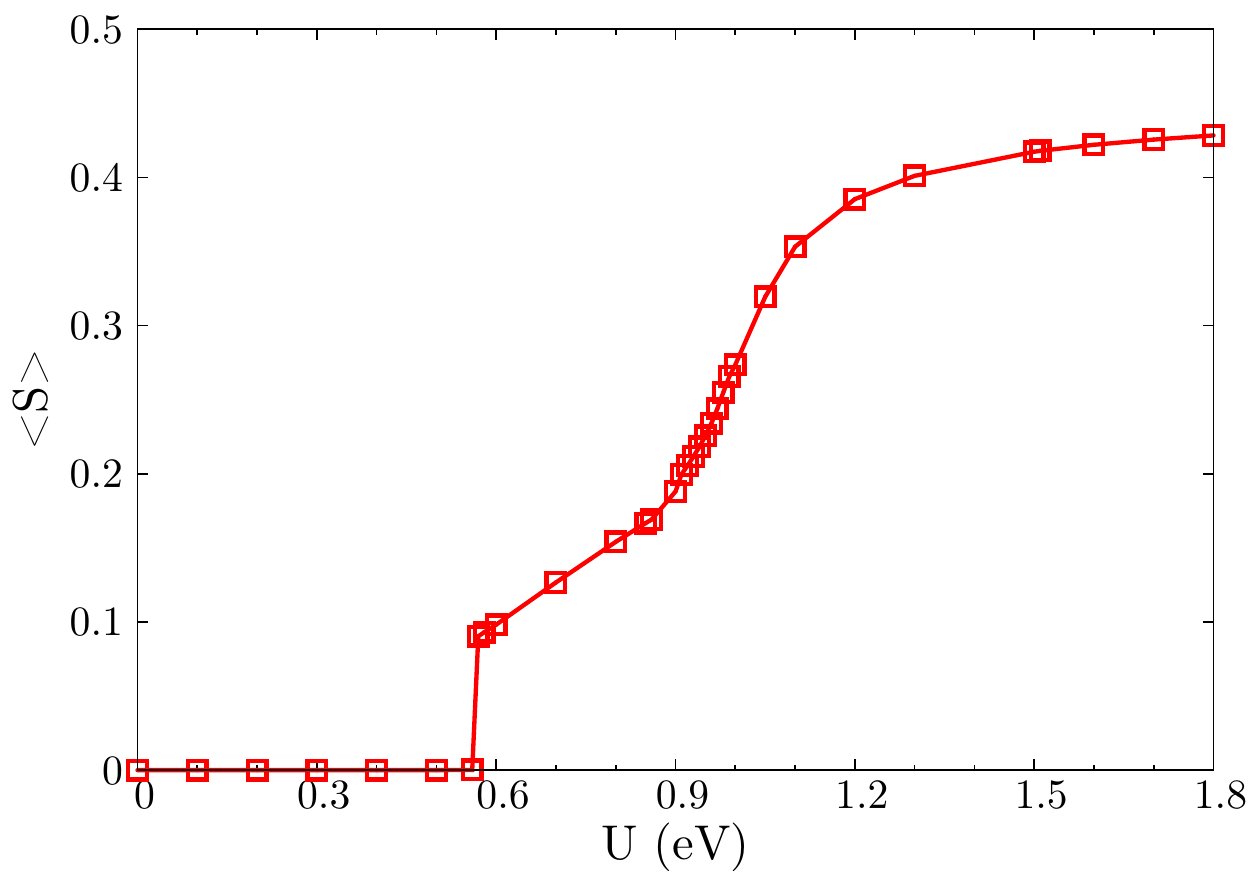}
\caption{Expectation value  of the on-site magnetic moment $\left\langle S \right\rangle$ averaged over the Ir sites as a function of $U$. All the magnetic states are antiferromagnetic with the AIAO magnetic ordering. The sudden change of the slope at around 0.9eV is a sign of the transition from Weyl metal to WSM.
}
\label{mag}
\end{center}
\end{figure}

Fig. ~\ref{mag} shows the evolution with correlation strength of the magnitude of the expectation value of on-site magnetic moment.  We see that the transition from paramagnetic metal to AIAO metal is characterized by a discontinuity in $\left\langle S \right\rangle$, so we identify this transition as first order. Around $U$=0.9eV a qualitative break in the slope of $\left\langle S \right\rangle$ vs $U$  is evident; we associate this with the change from Weyl metal to Weyl semimetal. Finally for $U\gtrsim$ 1eV the behavior becomes slower; this corresponds to the AF-I phase. 

It is of interest to analyse these results further, determining the evolution with $U$ of the locations of the Weyl points and of the associated energy scales. However the CDMFT calculations, while state of the art, are performed with an exact diagonalization method that approximates a continuous density of states by a finite set of delta functions and  are subject to uncertainties associated with the finite bath size in the impurity solver and with the need to introduce an artificial broadening to plot spectral functions. The interesting behavior therefore cannot be read directly off from the CDMFT results.   In the rest of this paper we fit the numerical results to the  $k\cdot p$ perturbation theory introduced in Ref.~\onlinecite{dft} and use the results of the fit to obtain more detailed insights into the WSM phase. 

\section{$k\cdot p$ theory \label{kdotp}}

The low energy physics of the CDMFT solution is described by quasiparticles moving in an effective band structure defined by correlations (which affect the real part of the self energy), crystal structure (which determines the underlying band theory) and magnetic order (which reconstructs the bands). In this section we interpret the CDMFT quasiparticle dispersions using the theoretical model presented in Ref.~\onlinecite{dft} based on a combination of symmetry analysis and $k\cdot p$ perturbation theory. Ref.~\onlinecite{dft} shows that one may write the quasiparticle Hamiltonian   near the $L$ point of the pyrochlore  Brillouin zone as a $2\times 2$ matrix in the space of relevant bands as
\begin{equation}
H_{eff}({\bf{q}}) = H_0 \mathbf{1} +\vec{H}\cdot\vec{\tau}
\label{kp-ham}
\end{equation}
with $\tau_{x,y,z}$ the usual Pauli matrices acting in band space and
\begin{eqnarray}
H_0&=&E_L+\frac{q_{in}^2}{2m_3}+\frac{q^{\star 2}}{2m_4}\tanh\left(\frac{q_z}{q^\star}\right)^2
\label{E0} \\
H_x&=&c_2 q_{in}^3 \sin 3\theta=c_2\left(-q_x^3+3q_y^2q_x\right)
\label{Ex}
\\
H_y&=&\beta q_z+c_1 q_{in}^3 \cos 3\theta=\beta q_z+c_1\left(q_y^3-3q_x^2q_y\right)
\label{Ey} 
\\
H_z&=&\Delta+\frac{q_{in}^2}{2m_1}+\frac{q_z^2}{2m_2}
\label{Ez}
\end{eqnarray}
Here $\bf{q}$ denotes the momentum measured relative to the $L$ point of the Brillouin zone, expressed in units of $2\pi$ divided by the basic (paramagnetic) pyrochlore lattice constant $a$. We  denote the $L\rightarrow\Gamma$ direction as $\hat{z}$ with the positive $q_z$ direction running from $L$ to $\Gamma$ and the projection in the perpendicular plane (the zone face $L$-$W$-$K$) as ${\bf{q}}_{in}$, with $\theta$ to be the angle between ${\bf{q}}_{in}$ and $L$-$K$ (see Fig. \ref{BZ} and \ref{LWK}). In this coordinate system $\Gamma$ is $(0,0,\sqrt{3}/2)$, $W$ is $(0.25\sqrt{2},0.25\sqrt{6},0)$, $K$ is $(0,0.25\sqrt{6},0)$ and the $y$-axis is $\theta=0$ and $x$ is $\theta=\frac{\pi}{2}$.   We have written the $q_z$-dependence of $H_0$ in terms of a function that as required by symmetry  is quadratic at small $q_z$ but constant at $q_z$ larger than a scale $q^\star$ because as we shall see $q^\star$ is comparable to the other scales in the problem. 
The eigenvalues of Eq.~\ref{kp-ham} are
\begin{equation}
E_\pm(q)=H_0(q)\pm\sqrt{\vec{H}(q)\cdot\vec{H}(q)}
\label{Edef}
\end{equation}

The electron Green function corresponding to this Hamiltonian is 
\begin{equation}
G(k,i\omega_n)=\sum_{s=\pm1}\frac{\mathbf{C}^{s}_k}{i\omega_n-E_s(k)}
\label{Gexplicit}
\end{equation}
with
\begin{equation}
\mathbf{C}^{s}_k=\frac{1}{2}\left(1+s\vec{h}_k\cdot\vec{\mathbf{\tau}}\right)
\label{Cdef}
\end{equation}
and $\vec{h}_k=\frac{\vec{H}(k)}{\left|\vec{H}(k)\right|}$.

The  band inversion parameter $\Delta$ is negative in the Weyl metal and semimetal phases; the transition to a trivial insulator is marked by a sign change in $\Delta$. The masses $m_{1\cdots4}$ and the constants $c_{1,2},\beta$ are to be determined by fits to the calculated quasiparticle band structure and are expected to depend weakly on $U$ in the vicinity of the critical value. 

\def\imagebox#1#2{\vtop to #1{\null\hbox{#2}\vfill}}

\begin{figure}
\begin{center} 
\hspace*{0.5em}
\subfigure[Brillouin zone]{
\label{BZ} 
\imagebox{43mm}{\includegraphics[width=0.42\linewidth]{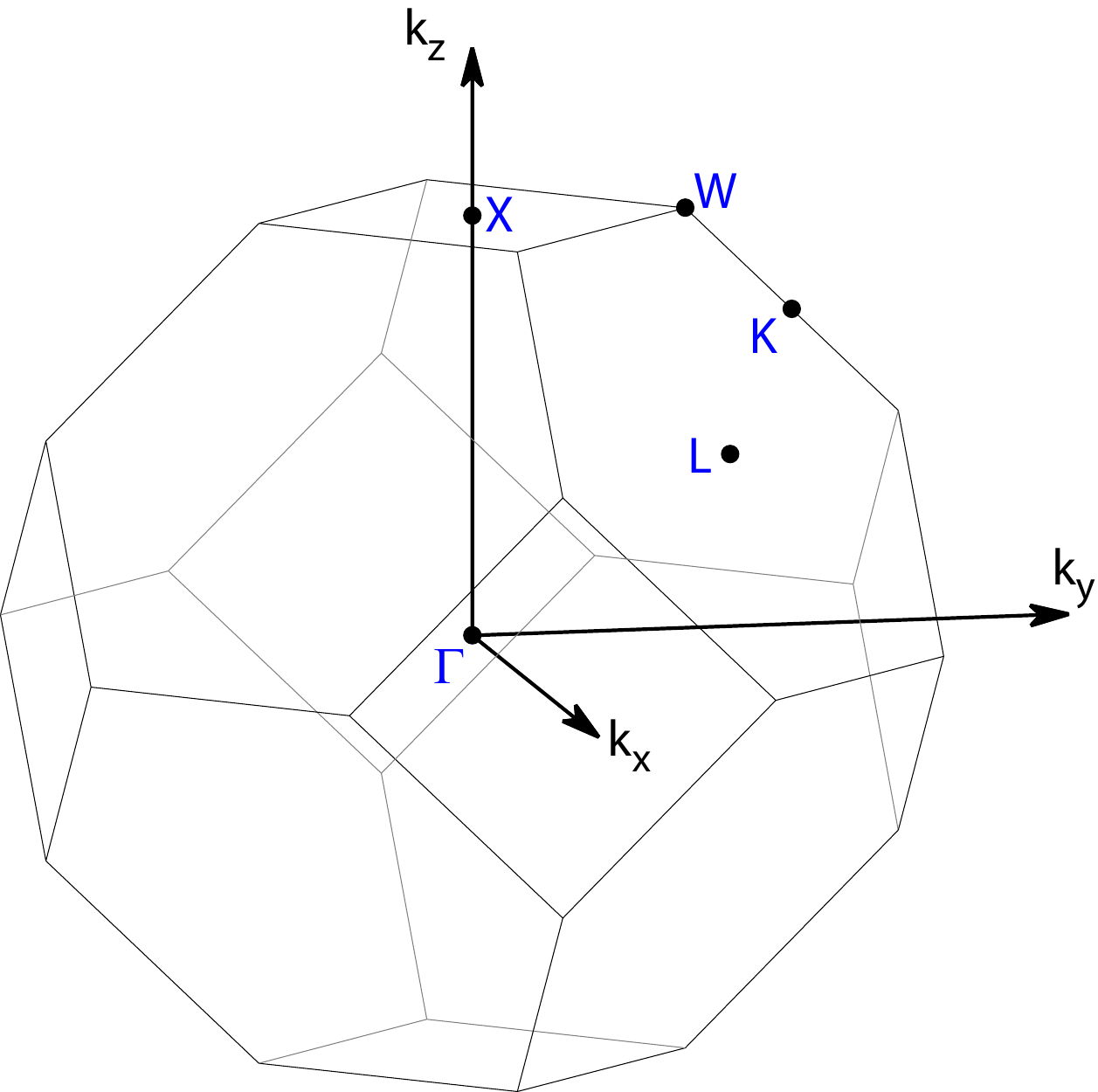}}}
\hspace*{1.5em}
\subfigure[$L$-$W$-$K$ zone face]{
\label{LWK} 
\imagebox{43mm}{\includegraphics[width=0.42\linewidth]{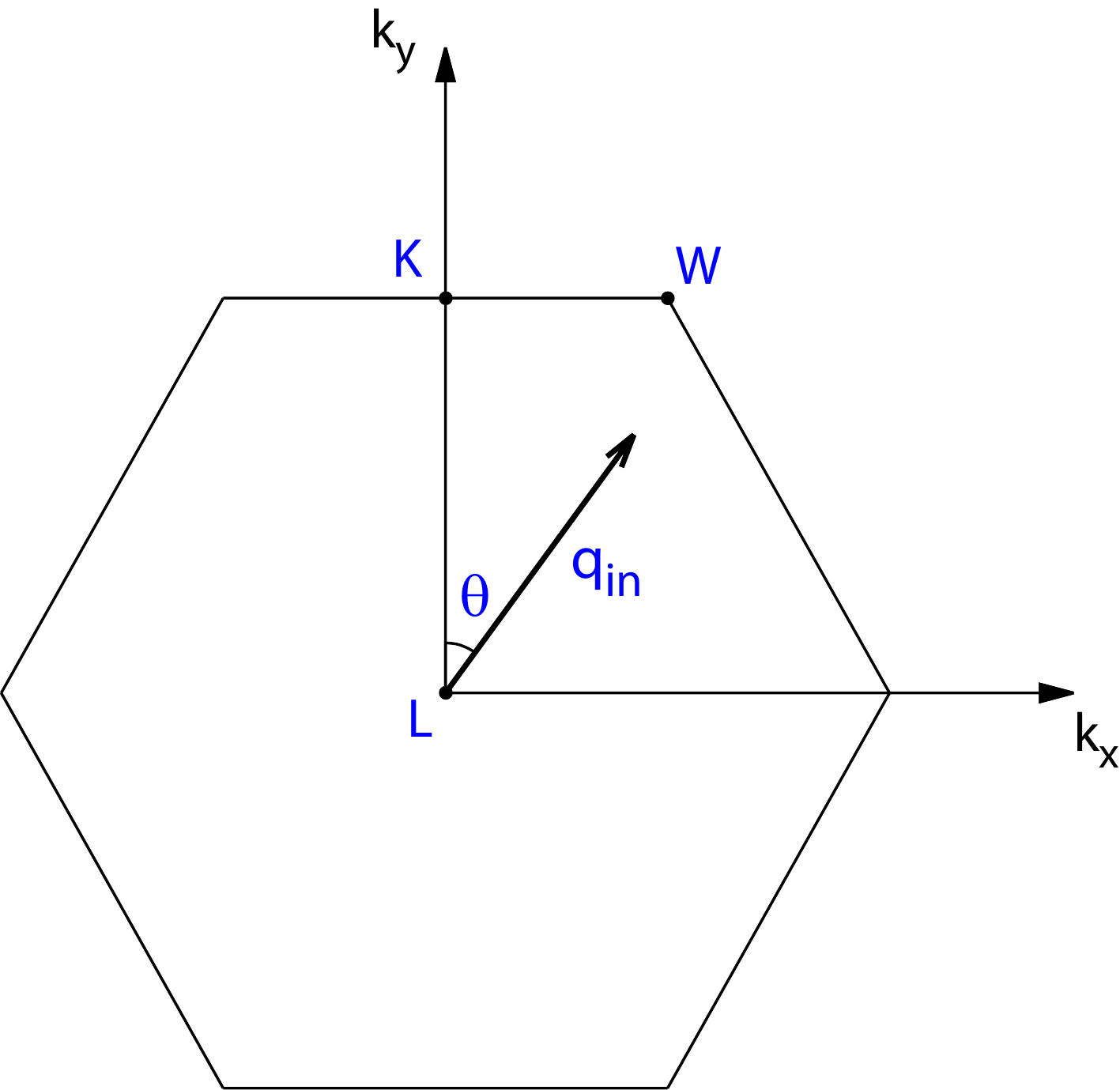}}}
\subfigure[$q_z$=0]{
\label{fs1}   
\imagebox{43mm}{\includegraphics[width=0.48\linewidth]{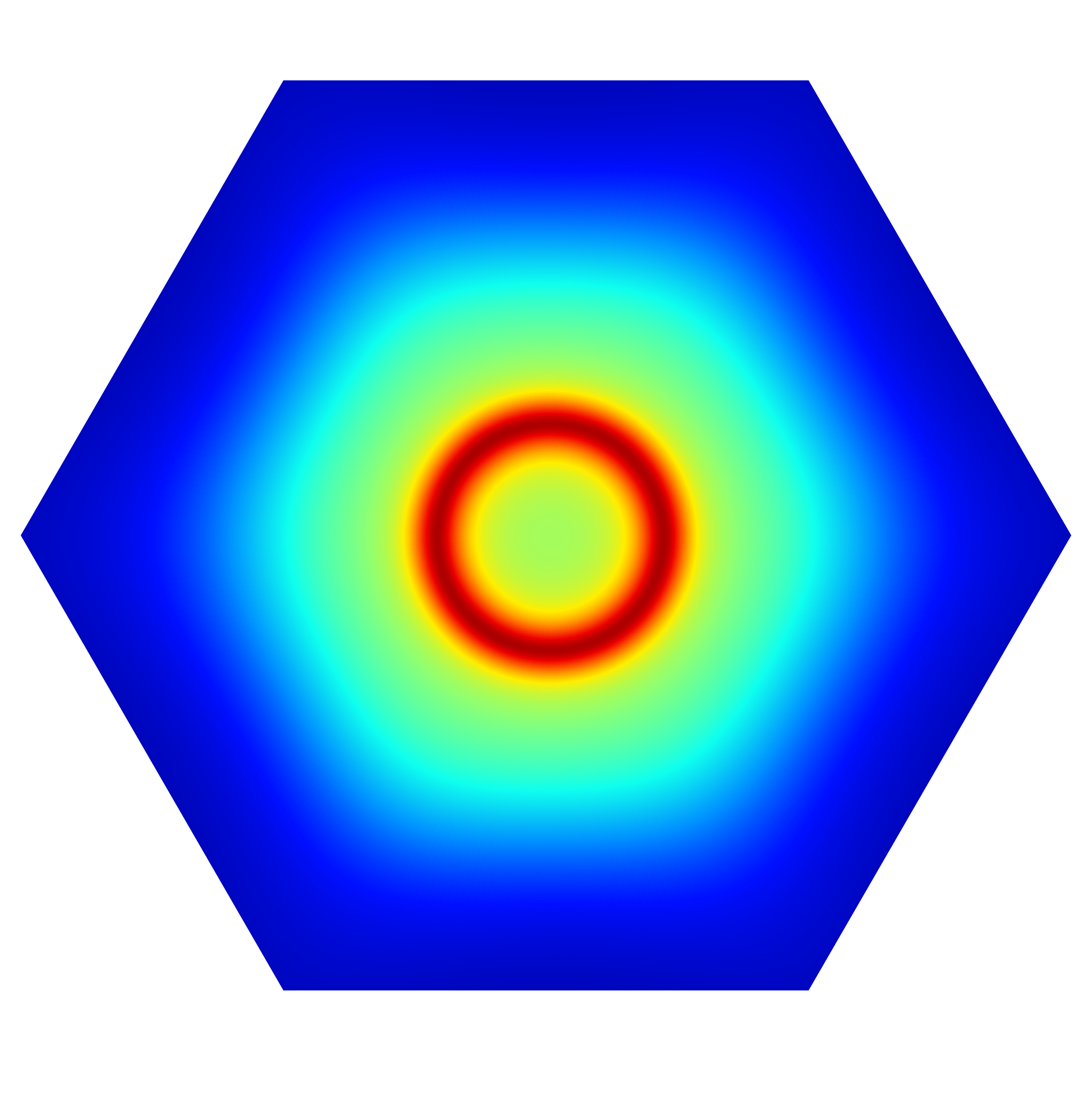}}}
\subfigure[$q_z$=0.0038]{
\label{fs2}  
\imagebox{43mm}{\includegraphics[width=0.48\linewidth]{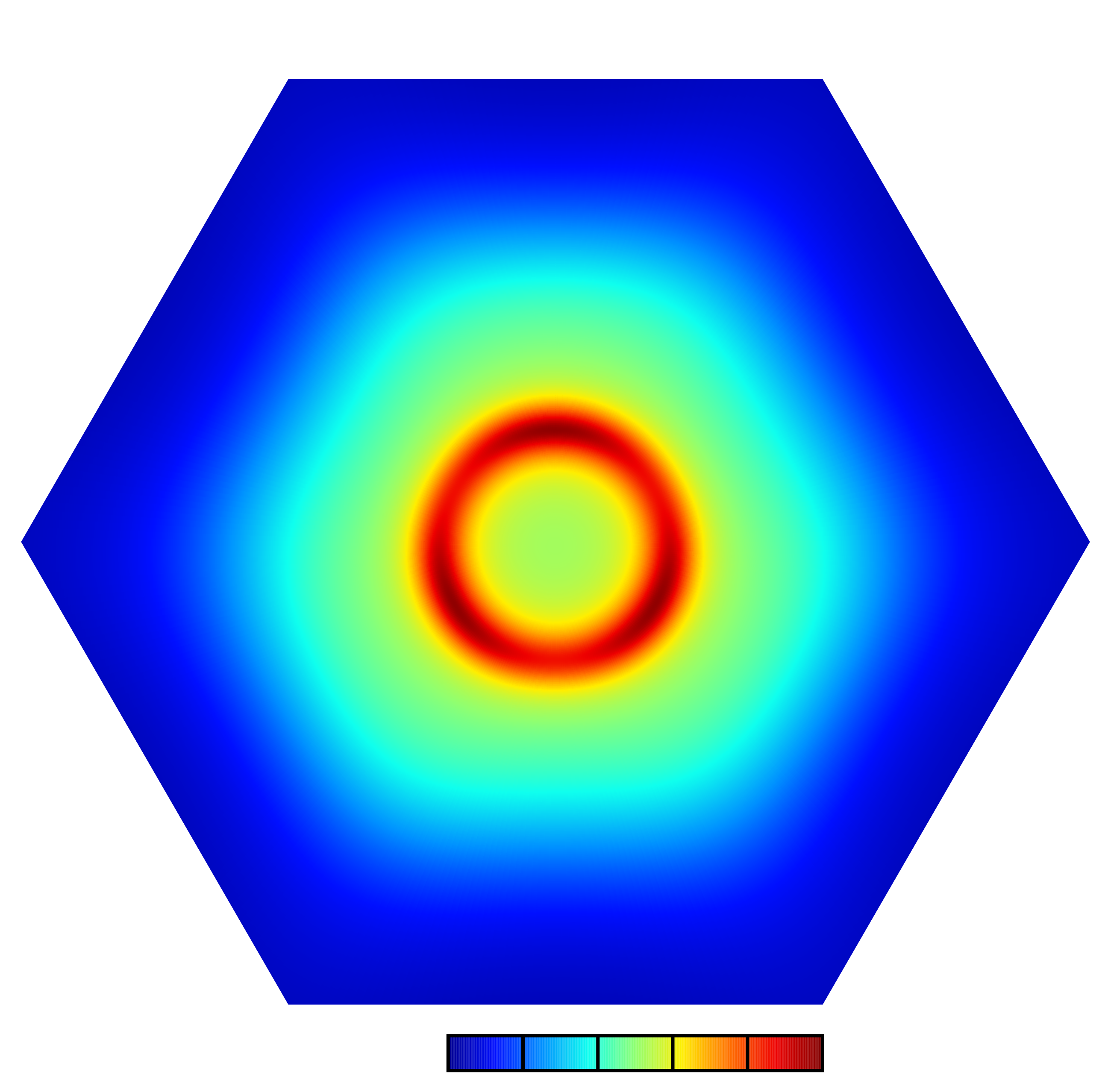}}}
\subfigure[spectral weight]{
\label{weight}   
\includegraphics[width=0.96\linewidth]{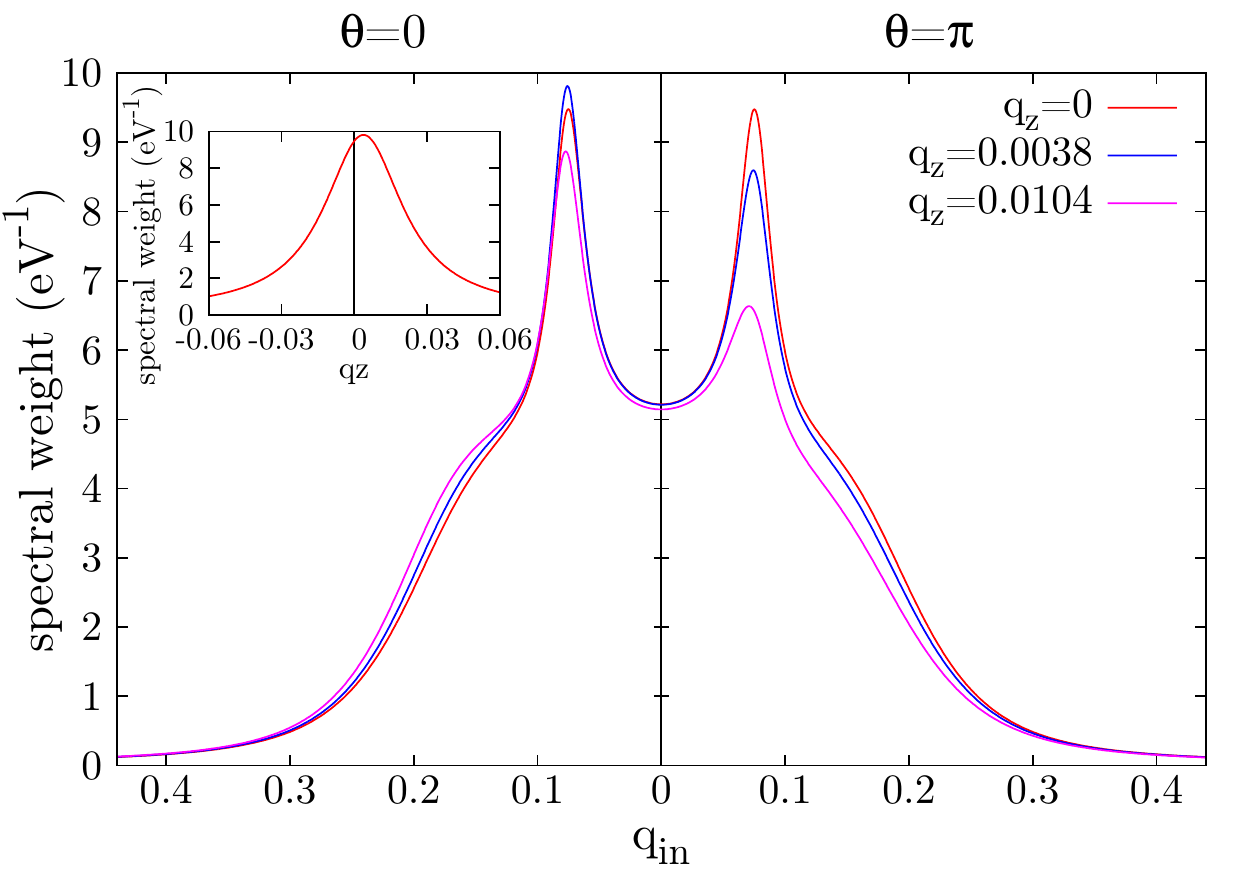}}
\caption{(a) Brillouin zone for the fcc lattice. (b) $L$-$W$-$K$ zone face and definition of angle $\theta$. (c) and (d): False color maps of spectral function (Eq.~\ref{Adef})  computed for $U$=0.96eV and broadening $\epsilon=0.005$eV at $\omega=0$ in  the $L$-$W$-$K$ plane (c) and the plane $q_z=0.0038$ containing the Weyl points (d). The center of the hexagon in panel (c) is $L$ (0.5,0.5,0.5), with $L\rightarrow K$ pointing in the upward direction (the orientation of the plane is the same as that shown in panel (b)). The plane shown in panel (d) is shifted from that in panel (c) along the $L\rightarrow\Gamma$ direction by an amount of 0.0038. The length of the sides for both hexagons shown here is half of that of the entire zone face $L$-$W$-$K$. (e) Spectral weight at the Fermi level for $U$=0.96eV plotted against in-plane momentum $q_{in}$ for two directions  ($\theta=0$, i.e. upward pointing and $\theta=\pi$ i.e. downaward pointing) in the $L$-$W$-$K$ plane or the parallel planes displaced along $q_z$.  The inset of this figure shows the spectral function at the Fermi level as a function of $q_z$ for $q_{in}=0.0760$ and $\theta=0$. The peak position is where one of the Weyl points sits.}  
\label{fs}
\end{center}
\end{figure}

The Hamiltonian involves many parameters but our numerical results indicate that some simplifications occur in the pyrchlore materials.  We first consider the zone face ($L$-$W$-$K$ plane, i.e. $q_z=0$) and focus on two directions, defined by $\theta=0$ and $\theta=\pi/2$. In this case we may write the eigenvalues of $H_{eff}$ (Eq.~\ref{kp-ham}) as

\begin{equation}
E_{\pm}(q_z=0,\theta=0)=E_L+\frac{q_{in}^2}{2m_3}
\pm\sqrt{
\left(\Delta+\frac{q_{in}^2}{2m_1}\right)^2+c_1^2q_{in}^6}
\end{equation}
\begin{equation}
E_{\pm}(q_z=0,\theta=\frac{\pi}{2})=E_L+\frac{q_{in}^2}{2m_3}
\pm\sqrt{
\left(\Delta+\frac{q_{in}^2}{2m_1}\right)^2+c_2^2q_{in}^6}
\end{equation}

\begin{figure}
\begin{center}
\includegraphics[width=1\columnwidth]{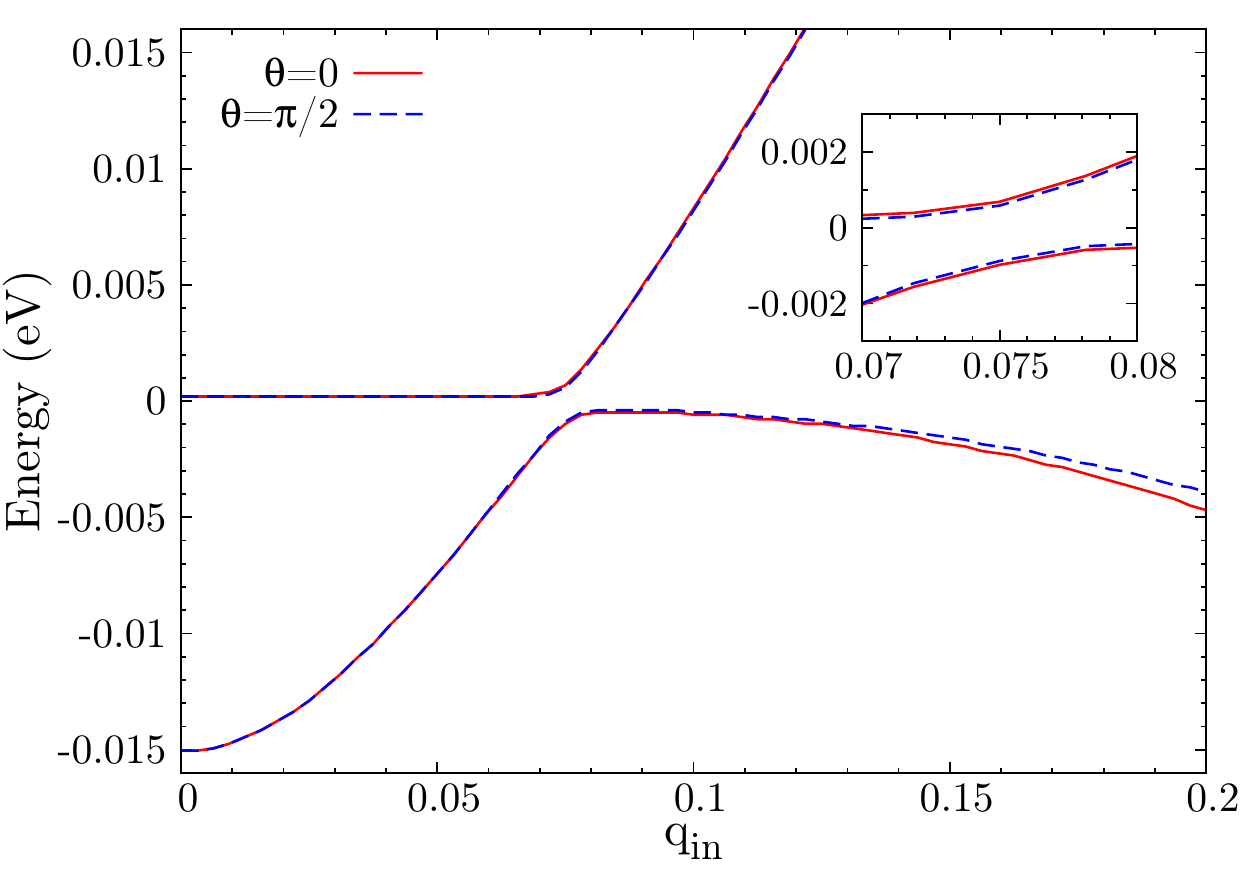}
\caption{The CDMFT band structure for the $k$-points on the $L$-$W$-$K$ plane ($q_z$=0) in the case of $U$=0.96eV. The red solid line is obtained for $\theta=0$ while the blue dashed line is obtained for $\theta=\pi/2$. The inset gives an expanded view of  the region of $q_{in}=0.07\sim0.08$ to highlight the approximately $\theta$-independent behavior for the $k$-points near $L$ on the $L$-$W$-$K$ plane.
}
\label{bands-qz0}
\end{center}
\end{figure}

Fig. \ref{fs1} shows a false color representation of the CDMFT spectral function (Eq.~\ref{Adef}) computed at $\omega=0$ in the $q_z=0$ plane. The line of maximum spectral weight forms an essentially perfect circle around the $L$ point.  To analyse the results  more quantitatively we have determined the quasiparticle bands (solutions of $det\left[G^{-1}(k,\omega)\right]=0$) by locating the peaks in $A(k,\omega)$. To find the peaks more precisely we used the smaller broadening $\epsilon=0.001eV$. Fig.~\ref{bands-qz0} plots the  resulting energies as a function of $q_{in}$ at $q_z=0$.  We see that the difference  $E(\theta=0)-E(\theta=\pi/2)$ is extremely small, even very near the crossing point. We conclude that we may set $c_1=c_2=c$. We also see that the upper band is approximately dispersionless for small $q$. This means $m_3\approx m_1$ (recall $\Delta<0$); we set these two masses equal henceforth. Next we see that the difference between the two eigenvalues is $\sim$0.015eV at $q_{in}=0$, from which we conclude that $\Delta\approx -0.0075$eV, while the difference between the two eigenvalues   almost vanishes at $q_{in}=0.075$, from which we conclude that $\frac{1}{2m_1}\approx 1.3$eV. Finally we observe that the minimum energy splitting is about 0.0017 $=2cq_{in}^3$ at $q_{in}=0.075$ so that $c\approx 2$eV.  

We now turn to $q_z\neq 0$. A Weyl crossing occurs when the coefficients of all three $\tau$ operators simultaneously vanish, requiring (in the notations above)  that  $\theta=2m\pi/3$ with $m$ an integer (the solutions corresponding to odd integer multiples of $\pi/3$ correspond to negative $q_z$, i.e. appear at the opposite zone face),   that  $\frac{q^2_{in}}{2m_1}+\frac{q_z^2}{2m_2}=-\Delta$ and that $q_z=-\frac{c}{\beta}q_{in}^3$. Fig. ~\ref{fs2} shows a false color representation of the dependence of the spectral function on in-plane momentum  at $\omega=0$ and $q_z=0.0038$. A clear $\cos3\theta$ variation is seen around the circle, consistent with the appearance of Weyl crossings. Fig. \ref{weight} shows this behavior in more detail. The main panel plots the $\omega=0$ spectral weight  as a function of $q_{in}$ for several values of $q_z$.  We see that the spectral function is largest at $q_z$=0.0038 and in the $\theta=0$ (upward-moving) direction. The inset shows the $q_z$ dependence at $q_{in}=0.076$ and $\theta=0$. Again a clear maximum is evident. These considerations enable us to locate the Weyl nodes for $U=0.96$eV at $q_z=0.0038$, $q_{in}=0.076$ and $\theta=0,2\pi/3,4\pi/3$. The $q_z$ position of the Weyl node  implies $\beta\approx-0.2$ and the factor of twenty ratio between the in-plane and out of plane wave vectors means that in the vicinity of the Weyl nodes we may neglect  $q^2_z/(2m_2)$ relative to $q_{in}^2/(2m_1)$.

Using the values of $\Delta$ and $\beta$ we then set $q_{in}$ equal to its value at the Weyl point and use the  $q_z$ dependence of the eigenvalues  to estimate $m_2$ and $m_4$.  From the difference in eigenvalues we find that $\frac{\left|\Delta\right|}{m_2}\lesssim 0.1\beta^2$ so that the $q_z^2/(2m_2)$ term can be neglected for relevant momenta. The $q_z$ dependence of the sum of the eigenvalues is very well fit to the form given in Eq.~\ref{E0}  with $\frac{1}{2m_4}=-0.235eV$ (note the negative sign) and $q^\star=0.1$. 
\begin{figure}
\begin{center}
\includegraphics[width=1\columnwidth]{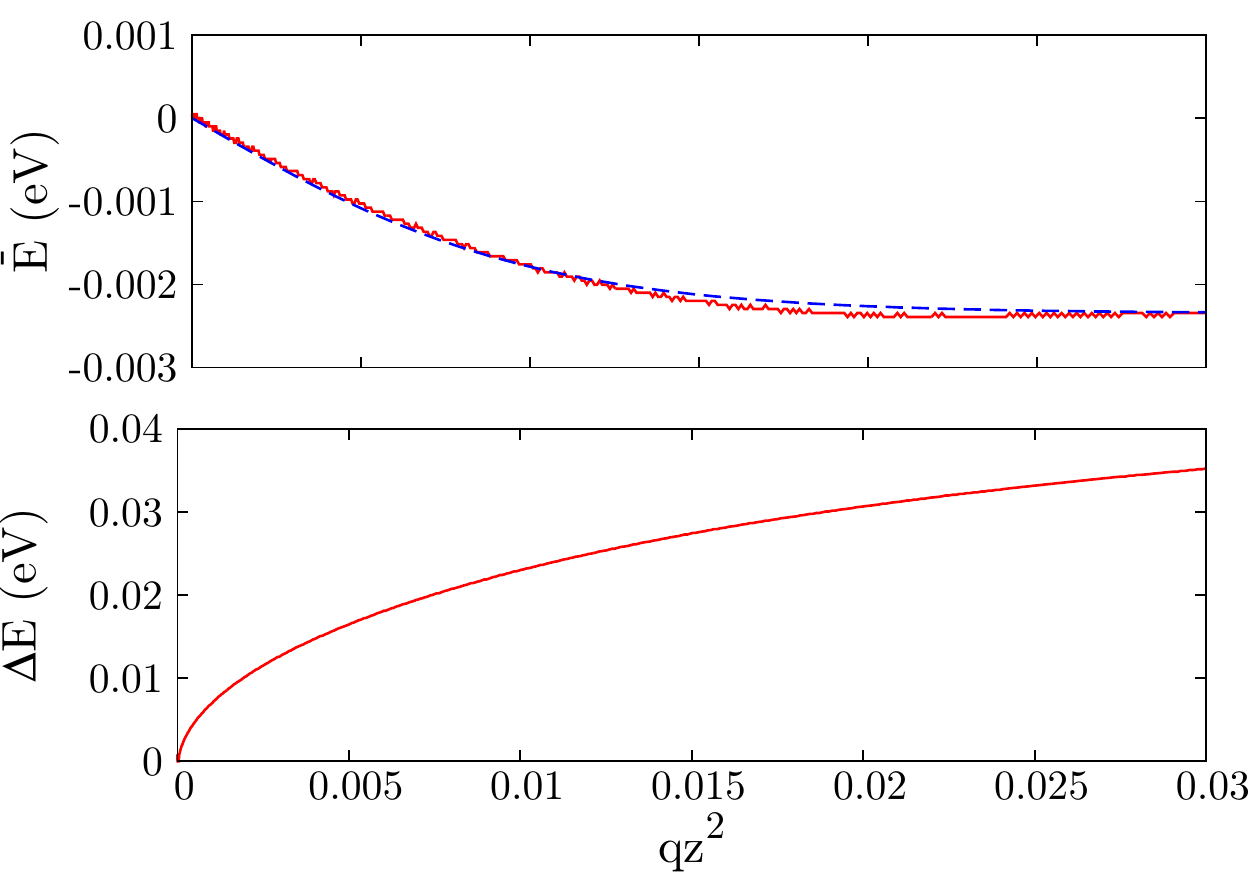}
\caption{The CDMFT band structure as a function of $q_z^2$ with $q_{in}$ to be the value of the Weyl point, in the case of $U$=0.96eV. Solid lines (red on-line): the top panel plots $\bar{E}=(E_++E_-)/2$ while the bottom panel plots $\Delta{E}=(E_+-E_-)/2$. The fitting to Eq. \ref{E0} is shown in dashed lines (blue on-line).
}
\label{fit-eigenvalue}
\end{center}
\end{figure}

\begin{table}
\caption{Quasiparticle band parameters for several $U$ values in the WSM regime, obtained as described in the text from fits to dispersions calculated with broadening $0.001eV$ except for $\frac{\beta}{c}$ which is obtained from the coordinates of the Weyl points at broadening $0.005eV$. Note that for $U=0.97$eV the very small value of the gap prevents a direct determination of $c$, and therefore $\beta$  from fits to the quasiparticle bands. $U$, $\Delta$, $\frac{1}{2m_1}$, $c$ and $\beta$ all have dimension of energy and are given in eV.}
\centering
\begin{tabular*}{\linewidth}{@{\extracolsep{\fill}}cccccccc}
\hline\hline
\hspace{0mm}U & $\Delta$& $\frac{1}{2m_1}$&   $c$ & $\beta$ &  $\frac{\beta}{c}$  \\[1ex]
\hline
\hspace{0mm}0.94 & -0.01284 & 1.569 & 2.375 & -0.2469 &  -0.1040 \\
\hspace{0mm}0.95 & -0.01045 & 1.425 & 2.178 & -0.2378 &  -0.1092 \\
\hspace{0mm}0.96 & -0.00762 & 1.345 & 1.947 & -0.2239 &  -0.1150 \\
\hspace{0mm} $0.97^*$ & -0.00430 & 1.277 &  &  &   -0.1127\\ [1ex]
\hline\hline
\end{tabular*}
\label{fit-para}
\end{table}

We have performed analogous fits of our numerical results for several other $U$-values. The results are summarized in Table ~\ref{fit-para}. We see that within our resolution $\Delta$ evolves smoothly and would change sign between $U=0.98$ and $0.99$eV while the other parameters remain non-critical, changing only slowly with $U$.

\section{Physical Consequences \label{Results}}

\subsection{Rescaled Hamiltonian}
The key result of the previous section is that the low energy  physics of the Weyl semimetal phase may be described as simplification of Eq.~\ref{kp-ham}  with one of the two hybridizing bands being essentially dispersionless in the $q_z=0$ plane.   In this section we derive some physical consequences of this somewhat unusual Hamiltonian. We also found that the angular anisotropy was very weak, so that to good approximation one has Weyl rings rather than Weyl points. Finally, the $q_z$ dependence of $H_z$ was found to be irrelevant at the energy scales of interest.  In this subsection we derive a rescaled Hamiltonian which displays the essential physics more clearly and then in subsequent subsections we present results for the low frequency optical conductivity and for susceptibilities. 

To begin our analysis we define a basic energy scale $E_0=\frac{1}{2m_1}=\frac{1}{2m_3}\approx 1.3eV$  and a momentum scale $q_W=\sqrt{-\Delta/E_0}$ and impose the equality $c_1=c_2$, obtaining for the energy eigenvalues (tildes denote energies normalized to $E_0$) \
\begin{eqnarray}
\tilde{E}_\pm&=&-\tilde{E_L}+q_{in}^2-\tilde{a}q^{\star 2}\tanh\left(\frac{q_z}{q^\star}\right)^2
\label{Epm}
\\
&\pm&\sqrt{\left(q_W^2-q_{in}^2\right)^2+\tilde{\beta}^2q_z^2+2\tilde{c}\tilde{\beta}q_zq_{in}^3\cos3\theta+\tilde{c}^2 q_{in}^6}
\nonumber
\end{eqnarray}
with $\tilde{c}\approx 1.4$,  $\tilde{a}\approx  0.17$ and $\tilde{\beta}\approx -0.17$. Here we have chosen the zero of energy to be  the Weyl crossing energy and we have retained the $q_zq_{in}^3$ and $q^6$ terms to regularize a divergence that will be found in the calculation of the susceptibilities. 

\begin{figure}
\begin{center}
\includegraphics[width=1\columnwidth]{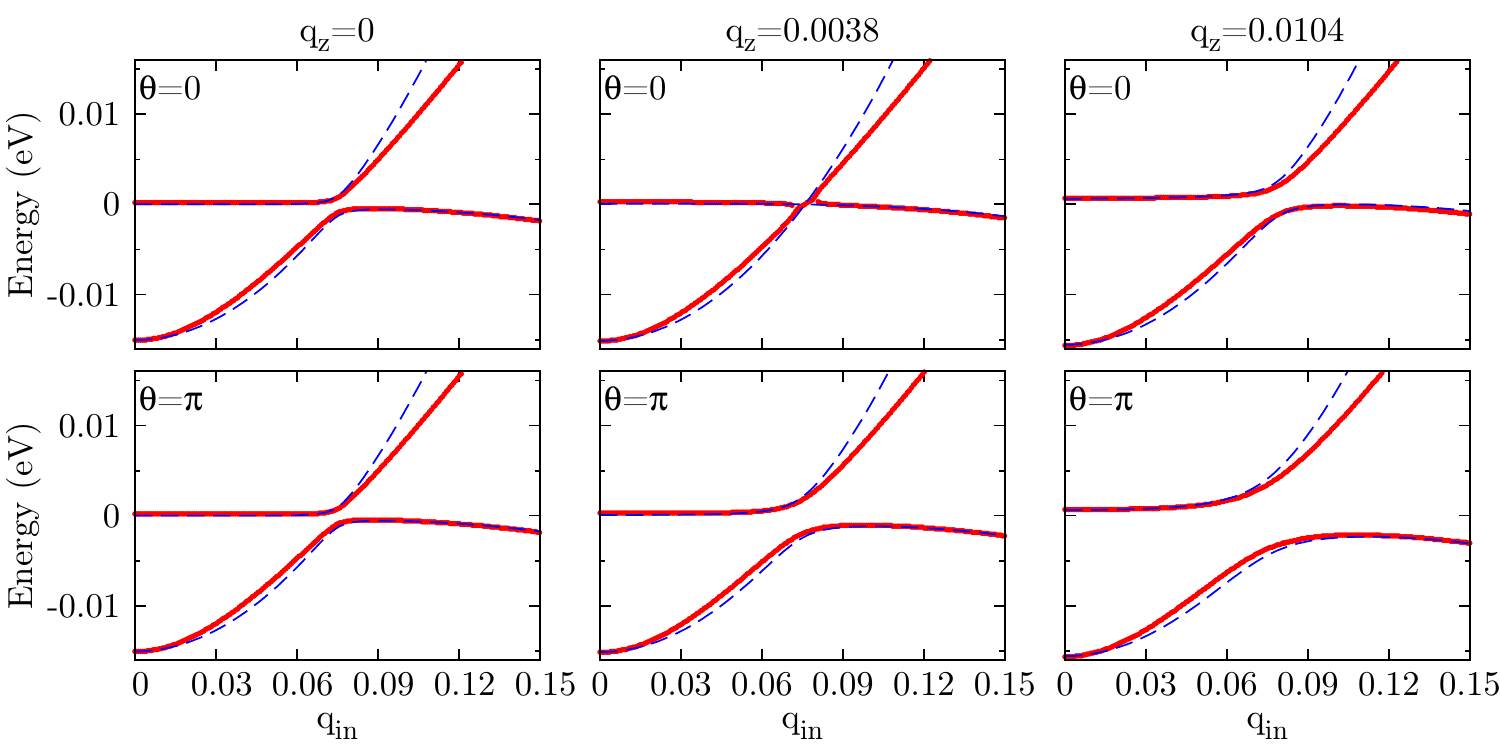}
\caption{Solid lines (red on-line): CDMFT band structure as a function of magnitude of  in-plane momentum  for several values of $q_z$ at $U$=0.96eV. Dashed lines (blue on-line): best fits to Eq.~\ref{kp-ham}.  }
\label{kp}
\end{center}
\end{figure}

Eq.~\ref{Epm} describes the hybridization of two bands with hybridization strength proportional to $q_z$ and $q_{in}^3$; for small $q_z$ one of the bands is nearly flat (dispersionless). The chemical potential is coincident with the bottom of the flat band. This behavior is clearly seen in Fig. ~\ref{kp}, which  plots the quasiparticle bands obtained from the CDMFT calculations, along with our best fits to the $k\cdot p$ expression, for several different values of $q_z$. 

The Weyl points are $q_{in}=q_W$, $\theta=0$ and $q_z=-\tilde{c}q_W^3/\tilde{\beta}$ (and symmetry-related points). The variation of the band gap  with $\theta$ for $q_{in}=q_W$, $q_z=-\tilde{c}q_W^3/\tilde{\beta}$ is $\Delta_W(\theta)=\Delta_W\left|\sin\frac{3\theta}{2}\right|$ with maximum gap 
\begin{equation}
\Delta_W=4\tilde{c}q_W^3
\label{Weylgap}
\end{equation}
smaller by one power of $q_W$ than the energy of the band minimum at the  $L$ point measured from the Weyl point. For $U=0.96$eV, the maximum $\Delta_W$ is about 15\% of this energy. 

If we are willing to neglect variations on the scale of $\Delta_W$ and focus on the region near the Weyl rings we may  neglect the  $q_z^2$ and $q_{in}^3$  terms in $\vec{H}$, obtaining

\begin{eqnarray}
\tilde{E}_\pm&=&-q_W^2+q_{in}^2-\tilde{a}q^{\star 2}\tanh\left(\frac{q_z}{q^\star}\right)^2
\nonumber 
\\
&&\pm\sqrt{\left(q_W^2-q_{in}^2\right)^2+\tilde{\beta}^2q_z^2}
\label{Epmring}
\end{eqnarray}
corresponding to the Hamiltonian
\begin{equation}
H_{simple}({\bf{q}_\parallel},q_z) = H_{simple,0} \mathbf{1} +\vec{H}_{simple}\cdot\vec{\tau}
\label{simple-ham}
\end{equation}
with $\tau_{x,y,z}$ the usual Pauli matrices acting in band space and
\begin{eqnarray}
H_{simple,0}&=&-q_W^2+q_{in}^2-\tilde{a}q^{\star 2}\tanh\left(\frac{q_z}{q^\star}\right)^2
\label{E0simple} \\
H_{simple,x}&=&0
\label{Hxsimple}
\\
H_{simple,y}&=&\tilde{\beta} q_z
\label{Hysimple} 
\\
H_{simple,z}&=&-q_W^2+q_{in}^2
\label{Hz}
\end{eqnarray}

\subsection{Optical Conductivity}

\subsubsection{Overview}

In this subsection we consider the optical conductivity.  Our analysis is similar to that of previous work  \cite{Nishine10,Ahn15,Tabert16,optical-conductivity}; the main differences arise from our focus on the  nearly flat band and  weak rotational symmetry breaking characteristic of the present case. 

In the quasiparticle approximation the dissipative part of the conductivity (a tensor in spatial indices giving the response to a translation-invariant, frequency-dependent electric field) is given by

\begin{eqnarray}
 \sigma^{ab}(\Omega)=\frac{1}{\Omega}  \frac{K^{ab}(\Omega+i0^+)-K^{ab}(\Omega-i0^+)}{2i}
\label{sigmadef}
\end{eqnarray}
with $K$ the analytic continuation of 
\begin{equation}
K^{ab}(i\Omega)=-Tr\left[\mathbf{J}^a(k)\mathbf{G}(k,i\omega_n+i\Omega)\mathbf{J}^b(k)\mathbf{G}(k,i\omega_n)\right]
\label{Kdef}
\end{equation}
where the current operators $J^{a=x,y,z}$ are hermitian matrices in band space that can be written
\begin{equation}
J^a=\vec{J}^a\cdot\vec{\tau}+J^a_0\mathbf{1}
\label{Jadef}
\end{equation}
and the trace is over momentum $k$, frequency $\omega$ and band indices.

By performing the standard quasiparticle computation and analytically continuing the result we obtain
\begin{equation}
\sigma^{ab}(\omega>0)=\frac{\pi}{\omega}\sum_k Tr\left[J^aC^+_kJ^bC^-_k\right]\delta\left(\omega-\delta E(k)\right)
\label{ImK}
\end{equation}
with $\delta E(k)=E_+(k)-E_-(k)$.

Noting that $C_\pm=\frac{1}{2}\left(1\pm\vec{h}\cdot\vec{\tau}\right)$ with $\vec{h}$ a unit vector and using standard Pauli matrix manipulations we find 
\begin{equation}
Tr\left[J^aC^+_kJ^bC^-_k\right]=\vec{J}^a\cdot\vec{J}^b-\left(\vec{J}^a\cdot\vec{h}\right)\left(\vec{J}^b\cdot\vec{h}\right)-i\left(\vec{J}^a\times\vec{J}^b\right)\cdot\vec{h}
\end{equation}

Note the appearance of an off-diagonal (Hall) term in the conductivity, arising because the Weyl point is a monopole in momentum space. 

The current operator is obtained from the fundamental quasiparticle Hamiltonian by making the Peierls substitution $\vec{k}\rightarrow \vec{k}-\vec{A}$ and differentiating the result with respect to $A$. From Eq.~\ref{kp-ham} we find (setting $c_1=c_2=\tilde{c}$, $m_1=m_3$ and using  the rescaled units defined in the previous subsection)

\begin{eqnarray}
J^x&=&2k_{in}\sin\theta\left(1+\tau_z\right)+3\tilde{c}k_{in}^2\left(\cos 2\theta\tau_x-\sin 2\theta\tau_y\right)\nonumber \\
 &\label{Jx}\\
J^y&=&2k_{in}\cos\theta\left(1+\tau_z\right)+3\tilde{c}k_{in}^2\left(\sin 2\theta\tau_x+\cos 2\theta\tau_y\right)
\nonumber \\
 &\label{Jy}\\
J^z&=&\tilde{\beta}\tau_y-2\frac{\tilde{a}k_z}{\cosh^2\left[\left(\frac{k_z}{q^\star}\right)^2\right]}
\label{Jz}
\end{eqnarray}
As a cross-check on Eqs. ~\ref{Jx}, \ref{Jy}, \ref{Jz} we calculated the current operator at the $L$ point in the standard band theory way (evaluating the matrix elements of  $\vec{\nabla}\cdot\vec{A}$ between the Wannier states and projecting the result onto $G^{-1}$) finding a result in agreement with Eqs. ~\ref{Jx}, \ref{Jy}, \ref{Jz}. In  particular this calculation confirms that there is a non-vanishing  matrix element between the two bands that become degenerate at the Weyl points.

Considering  the $\theta$ dependence of the current operators and noting that $\vec{h}$ has the full $C_3$ rotational symmetry about the  $L$-$\Gamma$ axis shows that the only nonzero components of $K$ are $K^{zz}$, $K^{xx}=K^{yy}$ and $K^{xy}=(K^{yx})^\star$.  Further, the $\delta$ function constrains $k_{in}^2$ to be of the order of the larger of $\omega$ and $q_W^2$, we may expand the current operator in powers of $k_{in}$. 

\subsubsection{Longitudinal conductivity}
For the diagonal components the leading terms are
\begin{equation}
\vec{J}^x=2k_{in}\sin\theta\hat{z},\hspace{0.1in}
\vec{J}^y=2k_{in}\cos\theta\hat{z},\hspace{0.1in}
\vec{J}^z=\tilde{\beta}\hat{y}
\label{Jleading}
\end{equation}
Eqs.~\ref{Jleading} then give
\begin{eqnarray}
\sigma^{xx}(\omega)&=&\frac{4\pi}{\omega}\int\frac{d^2k_{in}dk_z}{(2\pi)^3}k_{in}^2\sin^2\theta\frac{H_x^2+H_y^2}{\left|\vec{H}\right|^2}\delta\left(\omega-2\left|\vec{H}\right|\right)\nonumber\\
&\\
\sigma^{zz}(\omega)&=&\frac{\tilde{\beta}^2\pi}{\omega}\int\frac{d^2k_{in}dk_z}{(2\pi)^3}\frac{H_x^2+H_z^2}{\left|\vec{H}\right|^2}\delta\left(\omega-2\left|\vec{H}\right|\right)
\end{eqnarray}

The conductivities may be numerically evaluated by using the $\delta$ function to eliminate the $q_z$ integral and then performing the other two integrals numerically. Here we examine the two limits $\omega < \Delta_W$ and $\omega > \Delta_W$ where analytical results can be obtained. In the limit $\omega < \Delta_W$, the conductivity is dominated by the  Weyl points $k_{in}=q_W$, $k_z=k_z^\star=\pm \frac{\tilde{c}}{\tilde{\beta}}q_W^3$, $\theta=\theta_W= \frac{n\pi}{3}$ (n=0...5). Linearizing the Hamiltonian near any of the Weyl points gives
\begin{eqnarray}
H_x&=&3\tilde{c}q_W^3\cos3\theta_W\left(\theta-\theta_W\right)\equiv\delta_x
\\
H_y&=&\tilde{\beta}\left(k_z-k_z^\star\right)\equiv\delta_y
\\
H_z&=&2q_W( k_{in}-q_W)\equiv\delta_z
\end{eqnarray}
where we have defined local coordinates $\delta_{x,y,z}$.

For the diagonal terms in the conductivity we replace the $k_{in}$ in Eqs.~\ref{Jleading} by $q_W$, replace $dk_zk_{in}dk_{in}d\theta$ by $d^3\delta/(6\tilde{c}q_W^3|\tilde{\beta}|)$ and  sum over the 6 Weyl points,  obtaining
\begin{eqnarray}
\sigma^{xx}&=&\frac{2\pi q_W^2}{\tilde{c}q_W^3|\tilde{\beta}|\omega}\int \frac{d^3\delta}{(2\pi)^3} \frac{\delta_x^2+\delta_y^2}{\delta^2}\delta\left(\omega-2|\vec{\delta}|\right)
\nonumber
\\
&=&\frac{\omega}{12\pi \tilde{c}q_W|\tilde{\beta}|}
\\
\sigma^{zz}&=&\frac{\pi|\tilde{\beta}|}{\tilde{c}q_W^3\omega}\int \frac{d^3\delta}{(2\pi)^3} \frac{\delta_x^2+\delta_z^2}{\delta^2}\delta\left(\omega-2|\vec{\delta}|\right)
\nonumber
\\
&=&\frac{|\tilde{\beta}|\omega}{24\pi\tilde{c} q^3_W}
\end{eqnarray}

We now consider the regime $\omega>\Delta_W$ where we may make the Weyl ring approximation, replacing $\vec{H}$ by $\vec{H}_{simple}$ and performing the  integral over the in-plane angle we obtain
\begin{eqnarray}
\sigma^{xx}&=&\frac{2\pi}{\omega}\int \frac{k_{in}dk_{in}dk_z}{4\pi^2}k_{in}^2\frac{\tilde{\beta}^2k_z^2}{\left(q_W^2-k_{in}^2\right)^2+\tilde{\beta}^2k_z^2}
\nonumber \\
&&\times\delta\left(\omega-2\sqrt{\left(q_W^2-k_{in}^2\right)^2+\tilde{\beta}^2k_z^2}\right)
\\
\sigma^{zz}&=&\frac{\pi\tilde{\beta}^2}{\omega}\int \frac{k_{in}dk_{in}dk_z}{4\pi^2}\frac{\left(q_W^2-k_{in}^2\right)^2}{\left(q_W^2-k_{in}^2\right)^2+\tilde{\beta}^2k_z^2}
\nonumber \\
&&\times\delta\left(\omega-2\sqrt{\left(q_W^2-k_{in}^2\right)^2+\tilde{\beta}^2k_z^2}\right)
\end{eqnarray}

By performing the integration, we obtain the analytic form for the diagonal terms in the conductivity
\begin{widetext}
\begin{eqnarray}
\sigma^{xx}&=&\frac{q_W^2}{32|\tilde{\beta}|}\left\lbrace2+\Theta(\omega-2q_W^2)\left[-1+\frac{2}{3\pi q_W^2\omega^2}(\omega^2+2q_W^4)\sqrt{\omega^2-4q_W^4}-\frac{2}{\pi}\arctan\frac{-2q_W^2}{\sqrt{\omega^2-4q_W^4}}\right]\right\rbrace\\
\sigma^{zz}&=&\frac{|\tilde{\beta}|}{64}\left[2+\Theta(\omega-2q_W^2)\left(-1-\frac{4q_W^2}{\pi\omega^2}\sqrt{\omega^2-4q_W^4}-\frac{2}{\pi}\arctan\frac{-2q_W^2}{\sqrt{\omega^2-4q_W^4}}\right)\right]
\end{eqnarray}
\end{widetext}

\begin{figure}
\begin{center}
\includegraphics[width=1\columnwidth]{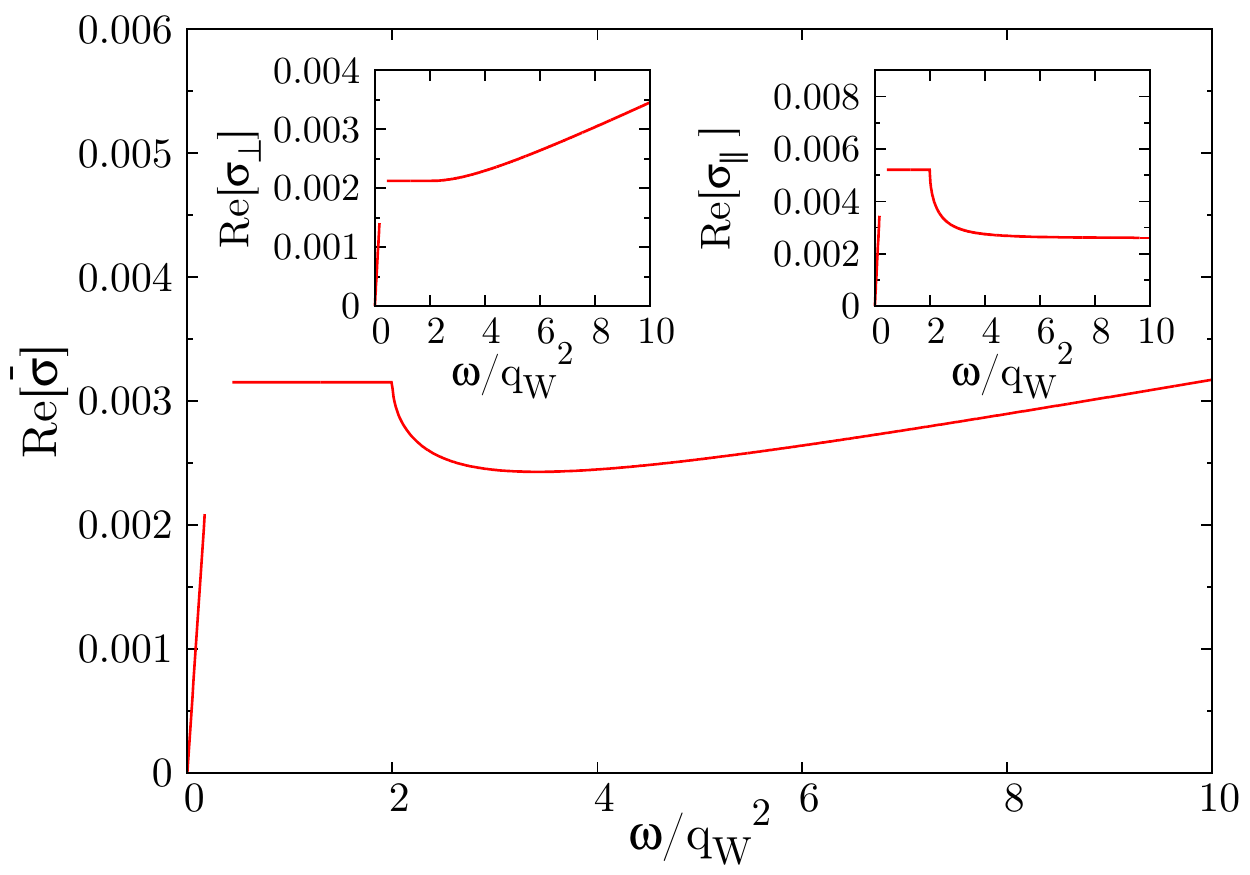}
\caption{Optical conductivity for $U$=0.96eV. The fitting parameters are chosen as demonstrated in Table \ref{fit-para}. We display the conductivity for $\omega <\Delta_W$ and $\omega>\Delta_W$; we have not obtained expressions for the crossover regime $\omega\sim \Delta_W$.}
\label{optical}
\end{center}
\end{figure}

The main panel of  Fig. \ref{optical} shows the observable conductivity $\bar{\sigma}\equiv(\sigma^{xx}+\sigma^{yy}+\sigma^{zz})/3$; the insets  plot $\sigma_{\bot}\equiv (\sigma^{xx}+\sigma^{yy})/2$ and $\sigma_{\parallel}\equiv\sigma^{zz}$. This structure is related to that found by Nishine, Kobayashi and Suzumura for a two dimensional system with a ``tilted Weyl cone" \cite{Nishine10}; see also the work of Ahn, Mele and Min \cite{Ahn15} for a related study of multi-Weyl semimetals, and Detassis et al for the conductivity associated with a tilted Weyl cone in three dimensions \cite{optical-conductivity}.  The optical conductivity of three dimensional  Weyl semimetals was also considered by Tabert and Carbotte \cite{Tabert16} (who obtained a different result because their study focussed on the case of symmetric bands (same velocity for electron and hole states)).  In the present case, the flat band/Weyl ring structure means  that the conductivity is frequency-independent for $\Delta_W<\omega<2q_W^2$ and has a square root singularity at    $\omega\sim 2q_W^2$. The nonanalaticity would be smoothed by terms of higher order in $q_W$ which we have neglected here.  We see that conductivity measurements (for our parameters, in the low THz or high GHz regime)  can reveal the basic energy scales of the Weyl semimetal. 

\subsubsection{Hall conductivity}

Working out the cross product needed for the off-diagonal term and noting that the Hamiltonian is even in $\tilde{\beta}k_z+\tilde{c}k_{in}^3\cos3\theta$  gives
\begin{equation}
\left(\vec{J}^x\times\vec{J}^y\right)\cdot\vec{H}=9\tilde{c}^2k_{in}^4\left(-q_W^2+k_{in}^2\right)-6\tilde{c^2}k_{in}^6\sin^23\theta
\label{crossproduct}
\end{equation}

We now consider the Weyl point-dominated regime $\omega<\Delta_W$. Inserting Eq.~\ref{crossproduct} in Eq.~\ref{ImK} and summing over the Weyl points gives

\begin{equation}
\sigma^{xy}=\frac{6\pi i}{\tilde{c}q_W^3|\tilde{\beta}|\omega}\int \frac{d^3\delta}{(2\pi)^3} \frac{\delta_x^2}{\left|\vec{\delta}\right|}\delta\left(\omega-2|\vec{\delta}|\right)=\frac{i\omega^2}{16\pi\tilde{c}q_W^3|\tilde{\beta}|}
\label{sigmaxylow}
\end{equation}

For  $\omega>\Delta_W$ we find, after performing the angle integration
\begin{eqnarray}
\sigma^{xy}&=&\frac{i\tilde{c}^2}{4\pi\omega}\int k_{in}dk_{in}dk_z \frac{9k_{in}^4q_W^2-6k_{in}^6}{\sqrt{(q_W^2-k_{in}^2)^2+\tilde{\beta}^2 k_z^2}}\nonumber\\
&&\times\delta\left(\omega-2\sqrt{(q_W^2-k_{in}^2)^2+\tilde{\beta}^2 k_z^2}\right)
\end{eqnarray}
Completing the integral we find
\begin{widetext}
\begin{eqnarray}
\sigma^{xy}(\omega>0)&=&\frac{3i\tilde{c}^2q_W^2}{128|\tilde{\beta}|\omega}\left\lbrace (8q_W^4-3\omega^2)\left[2+\Theta(\omega-2q_W^2)\left(-1-\frac{2}{\pi}\arctan\frac{-2q_W^2}{\sqrt{\omega^2-4q_W^4}}\right)\right]\right.\nonumber\\
&&-\left.\Theta(\omega-2q_W^2)\frac{4}{3\pi q_W^2}(2\omega^2-5q_W^4)\sqrt{\omega^2-4q_W^4}\right\rbrace
\label{sigmaxyhigh}
\end{eqnarray}
\end{widetext}

$\sigma^{xy}$ exhibits substantial structure, including a sign change at $\omega=\sqrt{8/3}q_W^2$ and a maximum $\sim q_W^3$ at $\omega\sim \Delta_W\sim q_W^3$, and at $\omega\sim\Delta_W$ the low and high frequency expressions are of the same order.

\subsection{Susceptibility}

We present here a qualitative discussion of the static real part of the  polarization functions (written here on the Matsubara axis)
\begin{equation}
\Pi^{ab}(q,i\nu)=Tr\left[\tau_a\mathbf{G}(k+q,i\omega+i\nu)\tau_b\mathbf{G}(k,i\omega)\right]
\label{Pidef}
\end{equation}
with $\mathbf{G}(k,\omega)=\left(\omega-\mathbf{H}(k)\right)^{-1}$, $\mathbf{H}$ is given by Eq. ~\ref{simple-ham} and the trace is over $k$, $\omega$ and band indices. A similar analysis focussing mainly on the two dimensional tilted Weyl case and emphasizing momentum space anisotropy was presented by Nishine et al \cite{Nishine10}, while various aspects of the three dimensional case were discussed by Fang, Chen, Kee and Fu   \cite{Fang15}. The focus here is on the consequences of the extreme flatness of one of the two bands that cross at the Weyl point and the very weak breaking of in-plane rotational isotropy. We consider mainly the $i\nu=0$ case, and restrict attention to in-plane $q$ ($q_z=0$).

Thus

\begin{equation}
\Pi^{ab}(q,i\nu)=T\sum_{nk;ss^\prime}\frac{Tr\left[\tau_a\mathbf{C}^s_{k+q}\tau_b\mathbf{C}^{s^\prime}_k\right]}{\left(i\omega_n+i\nu-E_s(k+q)\right)\left(i\omega_n-E_{s^\prime}(k)\right)}
\label{Pi2}
\end{equation}

Performing the sum over Matsubara frequencies, noting that we are dealing with one completely full and one completely empty band and that the energy is an even function of $k$ gives, at $\nu=0$,

\begin{equation}
\Pi^{ab}(q,\nu=0)=-\sum_{k}\frac{
Tr\left[\tau_aC^-_{k+q}\tau_bC^+_k+\tau_aC^+_{-k}\tau_bC^-_{-k-q}\right]}{E_{+}(k)-E_-(k+q)}
\label{Pi5}
\end{equation}

We analyse Eq. ~\ref{Pi5} by focussing on the singularities associated with regions where the denominator becomes very small, on the assumption (revisited below) that the numerator remains non-zero in the relevant range. We assume $q_z=0$.  From Fig.  \ref{kp}  we see that for $q=0$ the difference $E_-(k+q)-E_{+}(k)$ is of order $\Delta$ except along the Weyl ring, leading to a non-divergent $\Pi$ as we have already seen in the case of the optical conductivity. However for $q\neq 0$ there is a range of in-plane momementa  where $\left|k_{in}\right|<q_W$ and $\left|k_{in}+q\right|>q_W$ so both $E_+(k)$ and $E_-(k+q)$  are  small: in particular in this range at small $k_z$  we have   $E_+(k)-E_-(k+q)\sim \frac{\tilde{\beta}^2k_z^2}{2(q_W^2-k_{in}^2)}+\frac{\tilde{\beta}^2k_z^2}{2((k_{in}+q)^2-q_W^2)}$ so both terms are $\sim k_z^2$ and the $k_z$ integral diverges as $k_z^{-1}$  as $k_z\rightarrow 0$. For small  $|q|$, only the small  range where $\left|k_{in}\right|$ is close to $q_W$ and  $k_{in}$ and q are   in the same direction  exhibits this singular behavior, but as $|q|$  increases, the range of $k_{in}$ where the energy denominator is very small  grows  wider and for  $|q|>2q_W$ the energy denominator is very small for the entire range $\left|k_{in}\right|<q_W$.

The divergence at $k_z\rightarrow 0$  is cut off by the $k_{in}^3$ terms neglected in our simplified Hamiltonian Eq.~\ref{simple-ham}. A detailed analysis is very involved; here we note that Eq. ~\ref{Epm} shows that at $k_z=0$ the denominator in Eq. ~\ref{Pi5} is   $E_+(k)-E_-(k+q)\sim \frac{\tilde{c}^2k_{in}^6}{2(q_w^2-k_{in}^2)}+\frac{\tilde{c}^2(k_{in}+q)^6}{2((k_{in}+q)^2-q_W^2)}$. Combining this and what we get in the previous analysis for small $k_z$ and  considering  $|q|$ to be large so either $k_{in}<q_W$ and $|k_{in}+q|\sim q$ or $|k_{in}+q|<q_W$ and $|k_{in}|\sim |q|$, we estimate $E_+(k)-E_-(k+q)\sim \frac{\tilde{\beta}^2k_z^2}{2(q_W^2-k_{in}^2)}+\frac{\tilde{c}^2 q^4}{2}$. Rearranging,  we can estimate the polarizibility as 

\begin{eqnarray}
\Pi _{approx}&\sim&2\int^{q_W} \frac{d^2k_{in}}{(2\pi)^2} \int \frac{dk_z}{2\pi} \frac{2(q_W^2-k_{in}^2)}{\tilde{\beta}^2k_z^2+\tilde{c}^2 q^4(q_W^2-k_{in}^2)}
\nonumber
\\
&\sim& \frac{q_W^3}{3\pi q^2|\tilde{\beta}|\tilde{c}}\sim 0.1q_W
\end{eqnarray}
where in the final estimate we assumed $|q|=2q_w$ and used our numerical estimates for $|\tilde{\beta}|$ and $\tilde{c}$.  Thus if the numerator is non-vanishing  the susceptibility is of  order $q_W$.

Finally we consider the numerator
\begin{equation}
N^{ab}=Tr\left[\tau_aC^-_{k+q}\tau_bC^+_k+\tau_aC^+_{-k}\tau_bC^-_{-k-q}\right]
\label{Ndef}
\end{equation}
Because we are interested in  the divergence in $\Pi$ as $k_z\rightarrow 0$ and at $q_z=0$. we may  set $\vec{h}(k)$ in Eq.~\ref{Cdef} to be $\vec{h}(k)=\hat{z}sign(k_{in}^2-q_W^2)$. Because this is explicitly even in the sign of the momentum argument we may use the cyclic property of the trace and the fact that in the relevant region of integration $sign(k_{in}^2-q_W^2)=-1$ while $sign((k_{in}+q)^2-q_W^2)=1$  to obtain
\begin{equation}
N^{ab}=\frac{1}{2}Tr\left[\tau_a\left(1-\tau^z\right)\tau_b\left(1-\tau^z\right)\right]
\label{Ndef1}
\end{equation}

From this we immediately see that 
\begin{eqnarray}
N^{0,0}=N^{z,z}&=&2
\\
N^{0,z}=N^{z,0}&=&-2
\end{eqnarray}
and all the other components of $\mathbf{N}$ vanish at $k_z=0$ so these terms in the polarizibility are much smaller.  

Thus the flat bands mean that the susceptibilities are $\sim q_W$, parametrically larger than one would expect at a lightly doped three dimensional Dirac/Weyl point but probably not large enough to drive an instability. 

\section{{Conclusion} \label{Conclusion}}
This paper has analysed the transition from topologically nontrivial antiferromagnetic metal to topologically trivial antiferromagnetic insulator by interpreting and extending numerical results found in Ref. \cite{dft+cdmft} using the cluster dynamical mean-field theory of pyrochlore iridates in terms of a low-energy $k\cdot p$ approach. The mapping to a low energy theory was needed because the cluster dynamical mean-field theory results were obtained using an exact diagonalization solver. While the exact diagonalization solver has many advantages, including a direct computation of the real-frequency spectrum without recourse to analytic continuation, issues of bath discretization make it difficult to resolve  fine details of the spectrum. As we found, in the pyrochlore iridates  the Weyl semimetal phase is characterized by very small energy scales.  We therefore approach the problem by mapping the CDMFT results onto an analytic form suggested by $k\cdot p$ perturbation theory, which we then interpret as a quasiparticle Hamiltonian and study directly.

The  DFT+CDMFT results  clearly display three ground-state phases as the interaction strength is varied: a small $U$ paramagnetic, topologically trivial metal, an intermediate $U$ topologically nontrivial antiferromagnetic metal (``Weyl metal'') phase, and a topologically trivial insulator. The further analysis presented in this paper confirms that within the DFT+CDMFT method the Weyl metal and the antiferromagnetic  insulator phases are separated  by a Weyl semimetal phase existing over a narrow but non-infinitesimal range of $U$.  The transition to the Weyl semimetal state is marked by a change  in the variation with interaction strength of the total energy and of the magnetic moment. 

The transition from the Weyl metal to Weyl semimetal phase as $U$ is increased above the critical value $U_{WSM}$ occurs because an electron pocket centered at the $\Gamma$ point of the Brillouin zone gradually empties as the difference between the energy of the Weyl crossing point and the energy of the band minimum at the $\Gamma$ point gradually decreases. The transition from the Weyl semimetal to the topologically trivial insulator as $U$ is increased beyond the critical value $U_{AFI}$ occurs because the Weyl energy gap parameter $\Delta$ (Eq.~\ref{Ez}) gradually decreases in magnitude, passing through zero at the endpoint of the Weyl semimetal phase. In physical terms, as $U$ is increased the Weyl points move towards  high symmetry points (the $L$-points), where they annihilate. The existence of the Weyl semimetal phase requires that $U_{AFI}>U_{WSM}$; as far as we can see this condition is not enforced by any symmetry; the presence of the Weyl semimetal phase in the DFT+CDMFT calculation is from this point of view a particular feature of this theory of the pyrochlore iridates. 

It is also important to note that single-site dynamical mean-field theories find only the paramagnetic metal and antiferromagnetic insulator phases without the intermediate Weyl metal and Weyl semimetal phases. The striking difference between single-site and cluster dynamical mean-field results is remarkable in an electronically three dimensional compound. It is not yet clear whether the Weyl metal or Weyl semimetal phases are observed in pyrochlore iridates; while some indications have been found that data on the whole are most consistent with the presence only of antiferromagnetic insulator and paramagnetic  metal phases in these compounds. This difference between cluster DMFT and experiment remains to be understood.  However the Weyl semimetal phase is a clear prediction of the DFT+CDMFT theory of the pyrochlore iridates.

The Weyl semimetal phase found in the CDMFT calculations has two remarkable properties, both of which seem to be specific features of the model of the iridates rather than general consequences of symmetry. The first is the extreme weakness  of anisotropy in the plane of the zone face ($c_1\approx c_2$ in Eq.~\ref{kp-ham}).  This, combined with the symmetry-protected $q_z$-dependence, implies that to a very good approximation the material exhibits a ``Weyl ring'' (line in momentum space where the gap between the upper and lower bands is negligibly small). The second is that one of the two bands crossing at the Weyl point is essentially dispersionless in the $q_z=0$ plane. These two features lead to a  nonanalyticity in the optical conductivity and to an enhanced susceptibility. We also note that there is a Hall effect in the interband conductivity of the Weyl semimetal phase, arising from the topological properties of the Weyl point. Structure in the longitudinal and Hall conductivities is directly related to the energies of the Weyl semimetal phase, revealing both the energy parameter $\Delta$ and the scale $\Delta_W$  that characterizes the weak angular variation. More detailed investigation of these two features is an important task for future research.

\section*{Acknowledgement} 
RW and AJM acknowledge support by the Basic Energy Sciences Division of the DOE Office of Science under grant ER-046169. AG was supported by grant  IBS-R024-D1 from the Institute for Basic Science from the Ministry of Science, ICT, and Future Planning of Korea. The computing resources were provided by the Laboratory Computing Resource Center at Argonne National Laboratory.

\section*{{Appendix: Symmetrization}   \label{Appendix}}
The CMDFT calculations break lattice symmetries by including self energies for bonds within a cluster but not for symmetry-equivalent bonds connecting clusters. In the present case the basic CDMFT cluster is a 4-site tetrahedron which coincides with the unit cell of the pyrochlore lattice, so the CDMFT symmetry breaking arises because half of the symmetry-equivalent bonds under inversion operation are not included in our 4-site cluster. To restore the inversion symmetry, we need to symmetrize the results. We have carried out 2 symmetrization procedures: 1. symmetrizing the CDMFT self energy by including all the equivalent bonds; 2. symmetrizing the lattice green function in the momentum space. We find that both schemes can successfully restore the inversion symmetry and result only in very small corrections to most physical properties. However, the precise behavior of the Weyl crossing can be affected.  All the CDMFT results shown in the main text are computed by the first procedure (symmetrizing the self energy).
\subsection{Symmetrizing $\Sigma(\omega)$}  
In our one-orbital (per site) cluster model, the original CDMFT self energy is an $8\times8$ matrix composed of $2\times2$ blocks with site indices, which has the following form: 
\begin{equation}
\Sigma(\omega)
= 
\left[
\begin{array}{c|c|c|c}
\Sigma_{11} &\Sigma_{12} &\Sigma_{13} &\Sigma_{14} \\
\hline
\Sigma_{21} &\Sigma_{22} &\Sigma_{23} &\Sigma_{24} \\
\hline
\Sigma_{31} &\Sigma_{32} &\Sigma_{33} &\Sigma_{34} \\
\hline
\Sigma_{41} &\Sigma_{42} &\Sigma_{43} &\Sigma_{44}
\end{array}
\right]
\end{equation}
\\
According to this definition, we notice that half of the bonds, which are equivalent to those within our cluster under inversion, are not involved. To fix this, we apply the inversion operator to include all the nearest-neighbouring bonds. Therefore, for a given $\bf{k}$, the symmetrized self energy is calculated as follows: 
\begin{widetext} 
\begin{eqnarray}
\tilde{\Sigma}(\bf{k},\omega)
&=&\sum_{\bf{R}}\mathrm{e}^{i\bf{k}\bf{R}}\left\langle {\bf{0}}|\Sigma|{\bf{R}}\right\rangle \\
&=&\left[
\begin{array}{c|c|c|c}
\Sigma_{11} &
\frac{1}{2}(1+\mathrm{e}^{i\bf{k}\cdot (0,-1,0)})\Sigma_{12} &
\frac{1}{2}(1+\mathrm{e}^{i\bf{k}\cdot (0,0,-1)})\Sigma_{13} &
\frac{1}{2}(1+\mathrm{e}^{i\bf{k}\cdot (-1,0,0)})\Sigma_{14} \\
\hline
\frac{1}{2}(1+\mathrm{e}^{i\bf{k}\cdot (0,1,0)})\Sigma_{21} &
\Sigma_{22} &
\frac{1}{2}(1+\mathrm{e}^{i\bf{k}\cdot (0,1,-1)})\Sigma_{23} &
\frac{1}{2}(1+\mathrm{e}^{i\bf{k}\cdot (-1,1,0)})\Sigma_{24} \\
\hline
\frac{1}{2}(1+\mathrm{e}^{i\bf{k}\cdot (0,0,1)})\Sigma_{31} &
\frac{1}{2}(1+\mathrm{e}^{i\bf{k}\cdot (0,-1,1)})\Sigma_{32} &
\Sigma_{33} &
\frac{1}{2}(1+\mathrm{e}^{i\bf{k}\cdot (-1,0,1)})\Sigma_{34} \\
\hline
\frac{1}{2}(1+\mathrm{e}^{i\bf{k}\cdot (1,0,0)})\Sigma_{41} &
\frac{1}{2}(1+\mathrm{e}^{i\bf{k}\cdot (1,-1,0)})\Sigma_{42} &
\frac{1}{2}(1+\mathrm{e}^{i\bf{k}\cdot (1,0,-1)})\Sigma_{43} &
\Sigma_{44}
\nonumber
\end{array}
\right]
\label{sum}
\end{eqnarray}
\end{widetext}
where $\bf{R}$ is the real-space lattice vector, in the basis of Bravais lattice vectors of the fcc unit cell. A factor of $1/2$ has been introduced to the intersite terms which are doubly counted in the summation.

Based on our numerical tests, this symmetrization scheme works well in magnetic cases and paramagnetic metallic cases with small correction to the spectral gap and negligible correction to $\langle S \rangle$ and the energy. The scheme is problematic in the paramagnetic insulating case, as expected by analogy to the known issues with the standard self-energy  periodization scheme in CDMFT \cite{periodization}.

\subsection{Symmetrizing $G(\bf{k},\omega)$}
Motivated by the fact that the CDMFT Green function periodization can work for both metallic and insulating cases, we move to the symmetrization of $G(\bf{k},\omega)$. As required by the inversion symmetry, we simply replace the onsite blocks of $G(\bf{k},\omega)$ by the corresponding ones in $[G({\bf{k}},\omega)+G(-{\bf{k}},\omega)]/2$. This averaging scheme makes no difference to $\langle S \rangle$ and the energy due to a summation of $\bf{k}$ in the calculations. For the spectrum, it brings no correction to paramagnetic cases where time reversal symmetry is respected. On the other hand, in AIAO cases, the correction to the spectral gap turns out to be negligible according to our numerical tests.

\bibliography{Refs}

\end{document}